\documentclass[%
reprint,
amsmath,amssymb,
aps,
]{revtex4-2}

\usepackage{graphicx}
\usepackage{dcolumn}
\usepackage{bm}
\usepackage[mathlines]{lineno}
\usepackage{tabularx} 	
\usepackage{caption}	
\usepackage{rotating}	
\usepackage{subcaption}	
\usepackage{graphicx}	
\usepackage{datetime}
\newdate{date}{31}{10}{2019}
\date{\displaydate{date}}

\begin{document}
	
	\title{Aircraft Loading Optimization - QUBO models under multiple constraints}

	\author{Giovanni Pilon} \email{g.pilon@reply.it}
	\author{Nicola Gugole} \email{nicola\_gugole@outlook.it}
	\author{Nicola Massarenti} \email{nicola.massarenti@gmail.com}
	\vspace{5mm}
	\begin{abstract}
		\vspace{50mm}
		\textbf{In this submission we solve the Aircraft Loading Optimization problem of the Airbus Quantum Computing Challenge. Finding the optimal loading for a plane is a challenging task for classical algorithms, especially because the solution must respect several flight constraints. The contribution of this work is formulating this problem and its constraints in a model based on QUBO equations which are compatible with quantum annealers. We then benchmarked the model on different solvers to evaluate the performances and capabilities of current technologies.}
		\vspace{100mm}



	\end{abstract}


	\vspace{5mm}

	\maketitle

	\clearpage
	\newpage
	\mbox{~}
	\clearpage
	\newpage

\section{Introduction}

The objective of the Aircraft Loading Optimization is to maximize the payload of the aircraft while respecting several flight constraints \cite{airbus}. This problem can be viewed as an extension of the knapsack problem, a combinatorial optimization problem where the goal is to select the optimal set of items subject to a budget constraint. The knapsack problem can be written as a Quadratic Unconstrained Binary Optimization (QUBO) problem as showed in \cite{QUBO models}. 

QUBO Models are equivalent to the Ising problem formulation, and Quantum Annealers such as those developed by D-Wave Systems can be used as an optimization tool for these quadratic functions.

In this work we show that all the constraints of the Aircraft Loading Optimization Challenge proposed by \textit{Airbus} can be written in QUBO form, hence by treating them as penalty functions it is possible to assemble a QUBO model whose solution respects all the Aircraft Loading constraints. We then present the results obtained with a classical solver for QUBO functions and with a D-Wave Quantum Annealer accessed through API.

\section{Submission Summary}

In the section \textit{Detailed Explanation} we present the problem and the modelling for each constraint. In section \textit{Implementation} we explain the implementation and we present the used solvers. Finally in section \textit{Benchmarkig and performances} we present the obtained results for both the Quantum Annealer and the classical solver.
	
\section{Detailed Explanation}\label{Sec:DetailedExplanation}

As noted in the introduction, the goal is to maximize the aircraft payload subject to multiple constraints:
\begin{description}
	\item[PL - Payload Limits] The payload limits include conditions on the maximum capacity, the maximum allowed number of containers and their positions on the aircraft.
	\item[CL - Centre of Gravity Limits] The centre of gravity of the fully loaded aircraft must be within given bounds, moreover it must approach a given target value.
	\item[SL - Shear Limits] The shear of the fully loaded aircraft must be below a maximum shear curve.
\end{description}
We will handle the following types of containers:
\begin{description}
	\item[T1 - Medium Size] Requires a single position.
	\item[T2 - Small Size] Requires half of a single position.
	\item[T3 - Large Size] Requires two positions.
\end{description}
In the following, $n$ will denote the total number of available containers and $N$ the total number of available positions on the empty aircraft.
	\newcommand{\W}{\scriptscriptstyle{(W)}}
\newcommand{\N}{\scriptscriptstyle{(N)}}
\newcommand{\Obar}{\scriptscriptstyle{(\overline{O})}}
\newcommand{\C}{\scriptscriptstyle{(C)}}
\newcommand{\Dbar}{\scriptscriptstyle{(\overline{D})}}
\newcommand{\Ct}{\scriptscriptstyle{(C_t)}}
\newcommand{\Cl}{\scriptscriptstyle{(C_l)}}
\newcommand{\Cu}{\scriptscriptstyle{(C_u)}}
\newcommand{\Sl}{\scriptscriptstyle{(S_l)}}
\newcommand{\Sr}{\scriptscriptstyle{(S_r)}}

\subsection{Variables}\label{SubSec:Modeling}

We define the position assignment variables $p_{i,j}, \\ i=1,\dots, n, \ j=1,\dots,N$, such that 
\begin{equation}
	\label{eq:posvar}
	\begin{cases}
		p_{i,j} = 1, & \text{if container $i$ occupies position $j$}\\
		p_{i,j} = 0, & \text{otherwise}
	\end{cases} 
\end{equation}

We have a total of $nN$ binary target variables. This value does not include all the possible supplementary binary variables, the slack variables. In order to keep the notation simple, from now on $v_k$ will denote a generic slack variable and $c_k$ its coefficient.

\subsection{Writing the Problem in QUBO Form}
The goal is to write the problem in the QUBO form 
\begin{equation}
\label{eq:quboform}
\min_\mathbf{z} f^{(tot)} (\mathbf{z}), \quad f^{(tot)}(\mathbf{z}) \doteq \mathbf{z}^T Q \mathbf{z},
\end{equation}
where $Q$ is a symmetric square matrix of constants and $\mathbf{z}$ is a binary vector containing all the binary target variables introduced in \eqref{eq:posvar} and the supplementary slack variables. If we use the notation
\begin{align*}
	\mathbf{p} & \doteq (p_{i,j})_{i=1, \dots,n,  j=1,\dots,N} \\ 
	\mathbf{v} & \doteq (v_k)_k
\end{align*}
we can consider $\mathbf{z}$ as the concatenation of $\mathbf{p}$ and $\mathbf{v}$.

\subsection{Objective Function}
Each container $i$ has a given mass $m_i > 0$. Our objective function will be
\begin{equation*}
	f^{(obj)} (\mathbf{p}) = -\sum_{i=1}^n \sum_{j=1}^N t_i m_i  p_{i,j}
\end{equation*}
where
\begin{equation*}
	\begin{cases}
		t_i = 1, & \text{case T1 and T2}\\
		t_i = \frac{1}{2}, & \text{case T3}\\
	\end{cases} 
\end{equation*}
Since $f_{obj} (\mathbf{p})$ is linear in $\mathbf{p}$ it is compatible with the QUBO form of equation \eqref{eq:quboform}. Maximizing the loaded weight is equivalent to finding the minimum of the objective function. In the following sections we explain in detail how the constraints of the problem can be treated as penalties that influence the value of the objective function.

\subsection{Penalty functions}
The constraints of the problem are inequalities that must be respected by the solution. By default the QUBO formulation does not allow constraints, however, by rewriting these constraints as quadratic equations we can treat them as penalties that influence the value of the objective function (see \cite{QUBO models}). In particular these penalties are formulated in such a way that their value is zero for feasible solutions, and a positive amount for invalid solutions. The slack variables $v_k$ need to be introduced in order to transform inequalities into equalities (see section \ref{Sub:slacks}). Each penalty should be multiplied by a positive constant to have comparable magnitude with the objective function, we denote these constant with $P^{(-)}$.

\subsection{Payload Limits (PL)}

The payload limits (step 1) are composed of several constraints.

\vspace{2mm}
\noindent	
($\overline{O}$) \textit{No-overlaps}. The $p_{i,j}$ variables must represent a disposition of containers that is physically possible. The \textit{no-overlaps} condition imposes that each position is occupied at maximum by 1 medium/large container or by 2 small containers. For every position $j=1,\dots,N$, we impose
\begin{equation}
\label{eq:hvercons}
\begin{gathered}
\sum_{i=1}^n d_i p_{i,j} \leq 1, \\
\begin{cases}
d_i = 1, & \text{case T1 and T3}\\
d_i = \frac{1}{2}, & \text{case T2}
\end{cases}
\end{gathered}
\end{equation}
which can be translated into the penalty function
\begin{equation}
\label{eq:vfunc}
f^{\Obar}_j (\mathbf{p}, \mathbf{v}) = P^{\Obar} \Biggr(  \sum_{i=1}^n  d_i p_{i,j} +\sum_k c_k v_k- 1 \Biggl)^2.
\end{equation}

Equation \eqref{eq:vfunc} assumes positives values if an improper amount of container is assigned to the same position $j$.

\vspace{2mm}
\noindent
($\overline{D}$) \textit{No-duplicates}.
In the same way, each container must be assigned to only one position, except for large containers which must be assigned to 2 adjacent positions. We penalize every situation in which a container is assigned to a wrong amount of positions. To do this, for every container $i=1,\dots,n$ we impose
\begin{equation}
	\label{eq:horcons}
	t_i \sum_{j=1}^N  p_{i,j} \leq 1,
\end{equation}
which can be translated into the penalty function
\begin{equation}
	\label{eq:hfunc}
	f^{\Dbar}_i (\mathbf{p}, \mathbf{v}) = P^{\Dbar} \Biggr( t_i \sum_{j=1}^N  p_{i,j} + \sum_k c_k v_k - 1 \Biggl)^2, \quad P^{\Dbar} > 0.
\end{equation}

\vspace{2mm}
\noindent
($C$) \textit{Contiguity for big containers}.
Equation \eqref{eq:horcons} ensure that a large container occupies two positions but does not grant the contiguity of the two positions. For this reason, only for large containers, we extend equation \eqref{eq:horcons} by requiring that
\begin{equation}
\label{eq:horconsbig}
\begin{gathered}
	\sum_{j=1}^{N-1} (p_{i,j}-p_{i,j+1})^2 + p_{i,1} + p_{i,N} =  \sum_{j=1}^N  p_{i,j} \\
	\iff \sum_{j=1}^{N-1} p_{i,j}p_{i,j+1} = \frac{1}{2}\sum_{j=1}^{N} p_{i,j}.
\end{gathered}
\end{equation}
Indeed, if $\sum_{j=1}^{N} p_{i,j} = 2$ (large container case), we have
\begin{equation*}
	\sum_{j=1}^{N-1} p_{i,j}p_{i,j+1} = 1,
\end{equation*}
which is possible if and only if the selected positions are contiguous. Equation \eqref{eq:horconsbig} leads to
\begin{equation}
\label{eq:hbigfunc}
	f^{\C}_i (\mathbf{p}) = P^{\C} \Biggr( \frac{1}{2} \sum_{j=1}^{N} p_{i,j} - \sum_{j=1}^{N-1} p_{i,j}p_{i,j+1} \Biggl),
\end{equation}
$P^{\C} >0$. Equation \eqref{eq:horconsbig} already has degree two, for this reason it can't be treated like the other constraints (we can't raise it to power two to grant its positivity because it would generate a polynomial of order four, which is not compatible with the QUBO form). However, we can prove that the formulation of $f^{\C}( \mathbf{p})$ is already correct and consistent for proper values of $P^{\Dbar}$ and $P^{\C}$.

We need to prove that $f^{\Dbar}( \mathbf{p}) + f^{\C}( \mathbf{p})$ has positive value for unfeasible solutions and value zero for feasible solutions. From equations \eqref{eq:hfunc} and \eqref{eq:hbigfunc} we see that feasible solutions have a penalty of zero. It is trivial to see that $f^{\Dbar}( \mathbf{p}) + f^{\C}( \mathbf{p})$ is strictly greater than zero if $\sum_{j=1}^{N} p_{i,j} = 1$. We need to prove that the penalty value is strictly positive for the case $\sum_{j=1}^{N} p_{i,j} > 2$.

If we look at \eqref{eq:hbigfunc} we see that $f^{\C}_i (\mathbf{p})$ can be negative, specifically
\[
	f^{\C}_i (\mathbf{p}) < 0 \implies \sum_{j=1}^{N} p_{i,j} > 2.
\]
but if we set
\begin{equation*}
	P^{\Dbar} > 2 P^{\C}
\end{equation*}
we can guarantee that also in this case
\begin{equation*}
	f^{\Dbar}( \mathbf{p}) + f^{\C}( \mathbf{p}) > 0.
\end{equation*}

\textit{Proof.} If we substitute $\sum_{j=1}^{N} p_{i,j} = m$, then in the worst case $\sum_{j=1}^{N-1} p_{i,j}p_{i,j+1} = m-1$, which gives  
\begin{multline}
	f^{\Dbar}(\mathbf{p}) + f^{\C}(\mathbf{p}) \\
	= P^{\Dbar} \Biggl( \frac{1}{2} m -1 \Biggr)^2 + P^{\C} \Biggl( \frac{1}{2} m - (m-1) \Biggr) \\
	= \frac{1}{4} P^{\Dbar} (m-2)^2 - \frac{1}{2} P^{\C} (m-2)
\end{multline}
Then 
\[
	f^{\Dbar}(\mathbf{s}, \mathbf{p}) + f^{\C}(\mathbf{s}, \mathbf{p}) > 0 \iff P^{\Dbar} > \frac{2(m-2)}{(m-2)^2} P^{\C}
\]
The series 
\[
	\frac{2(m-2)}{(m-2)^2}, \quad m > 2
\]
has maximum at $m = 3$ and maximum value equal to $2$.

\vspace{2mm}
\noindent				
($W$) \textit{Maximum capacity}. 
In order to respect the maximum payload capacity of the aircraft $W_p$, we introduce the constraint
\begin{equation*}
	\sum_{i=1}^n \sum_{j=1}^N t_i m_i  p_{i,j} \leq W_p
\end{equation*}
which we transform into the penality function
\begin{equation*}
	f^{\W} (\mathbf{p}, \mathbf{v}) = P^{\W} \Biggr( \sum_{i=1}^n \sum_{j=1}^N t_i m_i  p_{i,j} + \sum_k c_k v_k - W_p \Biggl)^2
\end{equation*}
$P^{\W} >0$.

\subsection{Centre of Gravity Limits (CL)}
Given the length $L >0$ of the aircraft, we define the location of position $j=1,\dots,N$ as
\begin{equation*}
	x_j \doteq \frac{L}{N} \Biggl(j- \frac{N}{2} \Biggr) -\frac{L}{2N}.
\end{equation*}
Again, this step involves the use of multiple constraints.

\vspace{2mm}
\noindent
($C_t$) \textit{Center of gravity - target}.
We require that the actual center of gravity $\hat{x}_{cg}$ is equal to the target value $x^t_{cg}$ :
\begin{equation*}
	\begin{gathered}
			\hat{x}_{cg} \doteq \frac{\overline{x}_{cg}}{\underline{x}_{cg}} = \frac{\sum_{i=1}^n \sum_{j=1}^N t_i m_i p_{i,j} x_j + W_e x_{cg}^{e}} {\sum_{i=1}^n \sum_{j=1}^N t_i m_i  p_{i,j} +W_e}  = x^t_{cg} \\
			 \iff \overline{x}_{cg} = x_{cg}^t \underline{x}_{cg}
	\end{gathered}
\end{equation*}
where $W_e$ is the mass of the aircraft without payload, $x_{cg}^{e}$ is the corresponding center of gravity and $x_j$ is the horizontal coordinate for position j (see section \ref{sub:SL}). Then we obtain the following penalty function
\begin{equation}
	\label{eq:cgcons}
	f^{\Ct}(\mathbf{p}) = P^{\Ct} \Biggl(\overline{x}_{cg} - x_{cg}^t \underline{x}_{cg}\Biggr)^2, \quad P^{\Ct} > 0.
\end{equation}
Moreover, we want that the actual center of gravity is within the interval $[x_{cg}^{min}, x_{cg}^{max}]$ and we achieve this by adding the following penalty functions.

\vspace{2mm}
\noindent
($C_l$) \textit{Center of gravity - lower bound}.
\begin{equation}
	\label{eq:cgconsleft}
	f^{\Cl}(\mathbf{p}, \mathbf{v}) = P^{\Cl} \Biggl(\overline{x}_{cg} - x_{cg}^{min} \underline{x}_{cg}  - \sum_k c_k v_k  \Biggr)^2
\end{equation}
\vspace{2mm}
\noindent
($C_u$) \textit{Center of gravity - upper bound}.
\begin{equation}
	\label{eq:cgconsright}
	f^{\Cu}(\mathbf{p}, \mathbf{v}) = P^{\Cu} \Biggl(\overline{x}_{cg} - x_{cg}^{max} \underline{x}_{cg}  + \sum_k c_k v_k \Biggr)^2 
\end{equation}
where $P^{\Cl}, P^{\Cu} > 0$. Equations \eqref{eq:cgcons}-\eqref{eq:cgconsleft}-\eqref{eq:cgconsright} are compliant with equation \eqref{eq:quboform}. 

Notice that the value of $P^{\Cl}$ and  $P^{\Cu}$ should be higher than $P^{\Ct}$ because the upper and lower bound are hard constraints. However, by setting a high value for $P^{\Ct}$ it is possible to prioritize solution with an optimal center of gravity. In our implementation we choose $P^{\Cl} = P^{\Cu} \approx 10 P^{\Ct} $ to allow more flexible solutions.

\subsection{Shear Limits (SL)}
\label{sub:SL}

If we consider the symmetrical and linear case, then
\begin{equation*}
	S^{max} (x) \doteq 
	\begin{cases}
		S_0^{max}\Bigl( \frac{L+2x}{L} \Bigr), & x < 0\\
		S_0^{max}\Bigl( \frac{L-2x}{L} \Bigr), & x \geq 0
	\end{cases}, \quad  S_0^{max} >0.
\end{equation*}
Given the length $L >0$ of the aircraft, we define the location of position $u=1,\dots,N$ as
\begin{equation*}
	x_u \doteq \frac{L}{N} \Biggl(u-\frac{N}{2} \Biggr).
\end{equation*}
By discretizing the integral provided by \textit{Airbus} the actual shear force from the left at position $u$ is  
\begin{equation*}
	\hat{S}^l (u) \doteq 
	\sum_{i=1}^n \sum_{j =1}^{u} t_i m_i p_{i,j},  \quad u: x_u \leq 0
\end{equation*}
and the actual shear force from the right at position $u$ is 
\begin{equation*}
	\hat{S}^r (u) \doteq 
	\sum_{i=1}^n \sum_{j =u+1}^{N} t_i m_i p_{i,j}, \quad u: x_u \geq 0
\end{equation*}

In the following we consider the case where $N$ is even. 

\vspace{2mm}
\noindent
($S_l$) \textit{Shear - left}. We impose
\begin{equation}
\label{eq:sfevenconsl}
	\begin{gathered}
		\hat{S}^l (u) \leq S^{max} (x_u), \quad u=1,\dots,\frac{N}{2}\\
	\end{gathered}
\end{equation}
which correspond to summing, for $u=1,\dots,\frac{N}{2}$
\begin{equation*}
	f^{\Sl}_{x_u} (\mathbf{p}, \mathbf{v}) = P^{\Sl} \Biggr( \hat{S}^l (u) +\sum_k c_k v_k - S^{max} (x_u) \Biggl)^2.
\end{equation*}
where $P^{\Sl}>0$. 

\vspace{2mm}
\noindent
($S_r$) \textit{Shear - right}. We impose
\begin{equation}
\label{eq:sfevenconsr}
	\begin{gathered}
	\hat{S}^r (u) \leq S^{max} (x_u), \quad u=\frac{N}{2},\dots,N-1
	\end{gathered}
\end{equation}
which correspond to summing, for $u=\frac{N}{2},\dots,N-1$
\begin{equation*}
	f^{\Sr}_{x_u} (\mathbf{p}, \mathbf{v}) = P^{\Sr} \Biggr( \hat{S}^r (u) +\sum_k c_k v_k - S^{max} (x_u) \Biggl)^2
\end{equation*}
where $P^{\Sr}>0$. 

The case $N$ odd is similar; the main difference is that we have to consider the additional point $x=0$ as follows:
\begin{equation*}
	\begin{gathered}
		\hat{S}^l (u) \leq S^{max} (x_u), \quad u=1,\dots,\lfloor{\tfrac{N}{2}}\rfloor \\
		\hat{S}^r (u) \leq S^{max} (x_u), \quad u=\lfloor{\tfrac{N}{2}}\rfloor+1,\dots,N-1\\
		\hat{S}^l (\lfloor{\tfrac{N}{2}}\rfloor) +\sum_{i=1}^n \frac{t_i m_i p_{i,\lfloor{N/2}\rfloor+1}}{2} \leq S_0^{max}, \quad x=0 \\
		\sum_{i=1}^n \frac{t_i m_i p_{i,\lfloor{N/2}\rfloor + 1}}{2} + \hat{S}^r (\lfloor{\tfrac{N}{2}}\rfloor+1) \leq 	S_0^{max}, \quad x=0.
	\end{gathered}
\end{equation*}

Notice that this formulation is not bonded to the shape of the shear limit bound $S^{max}(x_u)$. In fact when writing equations \eqref{eq:sfevenconsl} and \eqref{eq:sfevenconsr} we don't need to make assumptions about the shape of $S^{max}(x_u)$ (in the implementation we treated $S^{max}(x_u)$ as an input). 

	
\subsection{Total function}
The total function is given by the sum of the object function and the penality functions, i.e., when $N$ is even:
\begin{align*}
	f^{(tot)} &= f^{(obj)} \\
	& + \underbrace{f^{\W} + \sum_{i=1}^n f^{\Dbar}_i + \sum_{i:i \text{ is T3}} f^{\C}_i + \sum_{j=1}^N f^{\Obar}_j}_{PL} \\
	& + \underbrace{f^{\Ct} + f^{\Cl} + f^{\Cu}}_{CL} \\
	& + \underbrace{\sum_{u=1}^{N/2} f^{\Sl}_{x_u} +\sum_{u=N/2}^{N-1} f^{\Sr}_{x_u} }_{SL}
\end{align*}
and similarly for the case in which $N$ is odd.
\subsection{Penalties scaling}
\label{Sub:penalties}
All the penalties should be opportunely scaled through their scaling coefficients $P^{(-)}$. In fact, the penalty functions and the objective functions should have the same magnitude. If a penalty function is order of magnitude lowers to the objective functions it becomes invisible for the solver, and the corresponding constraints will not be respected. Hence we calibrated the coefficients $P^{(-)}$ through sampling of the objective and penalty functions. The $P^{(-)}$ were increased until when the value of the penalties had at least the same magnitude of the objective function. It must be noted that the problem of calibrating the $P^{(-)}$ is crucial for a correct implementation, but at the same time it can be a very hard task if multiple constraints conflict one with the other.

\subsection{Heuristic method for slack variables}
\label{Sub:slacks}

Optimization problems with several constraints like this one need several slack variables. The method used to find the proper amount of slack variables must grant that the slack variables are enough but at the same time it should be parsimonious to avoid wasting resources (especially in this context where the problem dimension is a relevant issue).

A possible way to find the correct number of slack variables is to use a binary expansion method. Consider first a constraint in the form
\[
\boxed{\text{left side}} \leq u, \, u \geq 0 \Rightarrow \exists s_u: \boxed{\text{left side}} + s_u = u
\]
We have 
\begin{multline}
	0 \leq s_u = u-\boxed{\text{left side}}\\
	\leq \max \{u- \boxed{\text{left side}} \} = u - \min \{ \boxed{\text{left side}} \} \doteq \bar{u}
\end{multline}
There exists an integer $r_u$ and a realization of binary variables $\{ v_k \}$ such that
\[
\bar{u} \approx \sum_{k=0}^{r_u} 2^k v_k.
\]
Then we replace the constraint $\boxed{\text{left side}} \leq u$ by
\[
\boxed{\text{left side}} + \sum_{k=0}^{r_u} 2^k v_k = u.
\]
Similarly, a constraint of type $\boxed{\text{left side}} \geq l$ can be replaced by
\[
\boxed{\text{left side}} - \sum_{k=0}^{r_l} 2^k v_k = l.
\]

Although it is theoretically possible to fine tune the amount of slack variables to use for a given dataset, this method guarantees that all the constraint upper bounds and lower bounds are reached for any possible input dataset.

	\section{Implementation}

The model was coded in Python with the possibility of calling different solvers. Specifically we used:

\begin{itemize}
	\item{\textbf{\textit{QBsolv}}}: a classical solver for QUBO functions. \textit{QBsolv} is an open-source solver released by D-Wave Systems \cite{QBSOLV}, it is based on Tabu-Search, a metaheuristic optimization algorithm. Due to his heuristic nature this solver does not grant to reach the optimum of the problem, however it is able to find good sub-optimal solutions of medium-size problems ($ \sim 1000 $  variables) in a short amount of time. For this reason \textit{QBsolv} was useful to validate the model formalized in section \ref{Sec:DetailedExplanation}. We ran this solver locally on a normal laptop.
	
	\item{\textbf{D-Wave 2000Q}}: a Quantum Annealer accessed through D-Wave API. One of the main problems of a real Quantum Annealer hides in the mapping from a \textit{QUBO} function into a graph compatible with the hardware. In particular, the D-Wave quantum computer maps the \textit{QUBO} function into a graph, called \textit{Chimera Graph}. Our tests run on the lower noise D-Wave 2000Q (also called \textit{D-Wave 2000Q\_5}), with 2048 physical qbits. The \textit{Chimera Graph} of this quantum computer doesn't allow to map medium/big problems because the lattice of qubits isn't fully connected. To map the QUBO function to the \textit{Chimera Graph} we used the default heuristic created by D-Wave. Although it is possible to customize the embedding into the \textit{Chimera Graph} we did not explore this possibility. Being aware that the next generation of D-Wave annealers are expected to simplify the embedding we focused on the main challenges related to the modeling of the problem (the future Pegasus architecture will have over 5000 qbits, with a connectivity 2.5 times higher than the current one). Current anneling devices are still subject to various sources of imperfections and noise \cite{noise}, and in fact the obtained results present high variance and are often unfeasible solutions.
	
	\item \textbf{Exact solver}: a simple brute-force solver that finds the global optimum by testing all possible solution vectors. Since the time to brute-force the solution increases exponentially with the problem dimension we used this solver for very small instances of the problem. This solver does not actually brute-force through the QUBO model, but just through all the possible position assignment vectors \textbf{p}, searching for solution optimality and feasibility. In this way we reduce the computation time (the QUBO model is much harder to brute-force because it includes several slack variables).
	
\end{itemize}
	\section{Benchmarking and performances}
	
	This section shows the results obtained for the different solvers and different problem configurations. Section \ref{Sub:LocalSolver} presents examples of solutions obatained with the classical solver.  Section \ref{Sub:500} show the statistical performance of the classical solver over 500 test. Finally section \ref{Sub:DWave} shows an analysis of the performance on a real Quantum Annealer. For both the solvers we used their default settings. 
	
	\subsection{Classical solver}\label{Sub:LocalSolver}
	Testing the performance of the implementation by means of a classical solver allows to verify the correctness of the model formalized in Section \ref{SubSec:Modeling}. 
	
	The input data are shown in Tables \ref{Table:ContainerData} and \ref{Table:OtherData}, describing the constraints and the containers specifics in compliance with the indications provided by \textit{Airbus}, we added 5 large containers in the container dataset to prove that the model was capable of handling them correctly. For analyzing the performance of the model we developed several experiments involving a combination of the constraints \textit{PL}, \textit{CL} and \textit{SL}, defined in Section \ref{Sec:DetailedExplanation}.
		
		
		
		\subsubsection{Examples}
		\label{Sub:Examples}
		
\begin{figure*}[!htb]
	\centering
	\caption{Result of the computation with the local solver \textit{Qbsolv} with $n=35$, $N=20$, with parameters of Tables \ref{Table:ContainerData} and \ref{Table:OtherData} and constraint \textit{PL}.}
	\label{Fig:QbsolvPL}
	\begin{subfigure}[b]{0.6\textwidth}\centering
		\includegraphics[width=\textwidth]{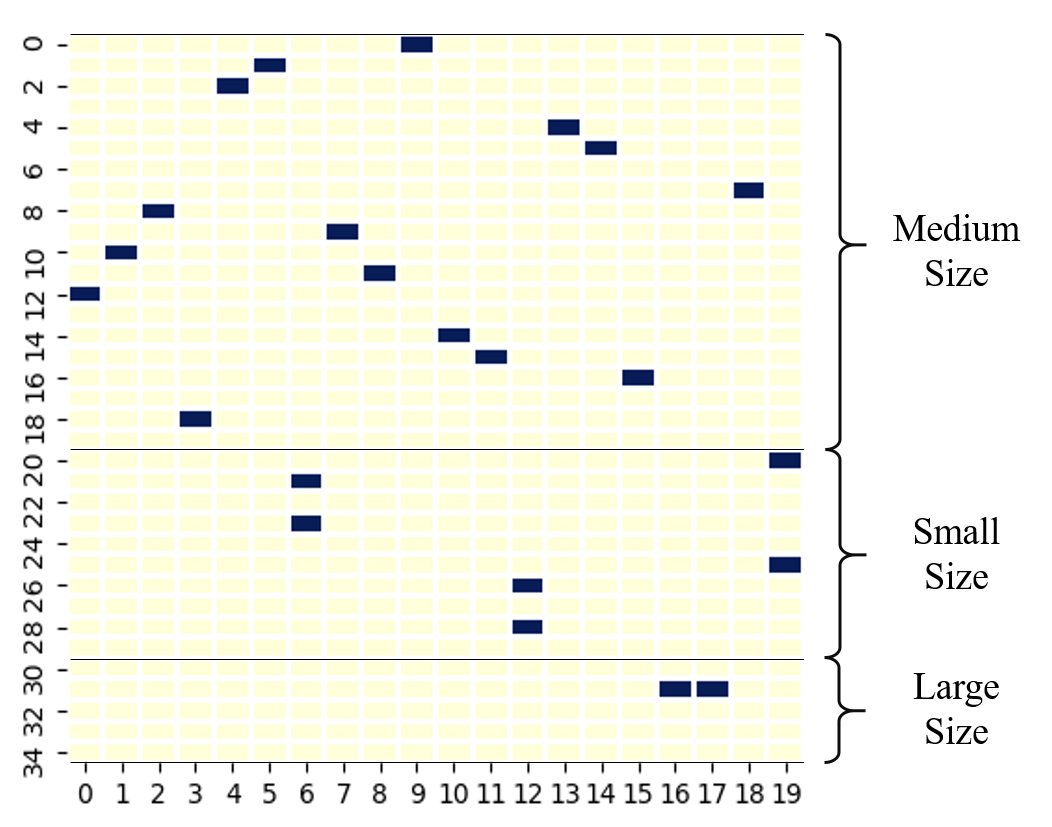} 
		\caption{Output configuration obtained with the local solver \textit{QBsolv}, using all the PL constraints except for $f^{\W}$. The abscissa represents the available slots on the aircraft, the ordinate represents the containers, hence position $j$ contains container $i$ if cell $p_{i,j}$ of the matrix is colored in blue. As it can be observed, each position assignment is correct: the lines reserved to medium containers have only one colored cell both in the horizontal and vertical direction, the lines reserved to small containers have only one colored cell horizontally and maximum two vertically, those reserved to big containers have exactly two colored cells horizontally and exacly one vertically.}
		\label{Fig:QbsolvPLOut}				
	\end{subfigure}	
	\quad
	\begin{subfigure}[b]{0.8\textwidth}\centering
		\includegraphics[width=\textwidth]{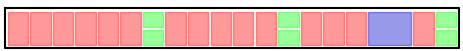}
		\caption{Disposition of the containers of the cargo using all the PL constraints except for $f^{\W}$: all the available positions are occupied, with a total resulting mass of $47739$ kg. The red boxes are the containers of type \textit{T1}, the green boxes are the containers of type \textit{T2} and the blue boxes are the containers of type \textit{T3}.}
		\label{Fig:QbsolvPLDisp_noW}
	\end{subfigure}
	\quad
	\begin{subfigure}[b]{0.8\textwidth}\centering
		\includegraphics[width=\textwidth]{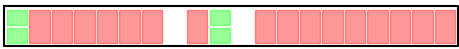}
		
		\caption{Disposition of the containers of the cargo, using all the PL constraints, with a total resulting mass of $39616$ kg. There are empty spaces because the maximum weight is almost saturated ($W_{max}=40000 $ kg).}
		\label{Fig:QbsolvPLDisp}
	\end{subfigure}		
	
\end{figure*}

\begin{figure*}[!htb]
\quad
\caption{Disposition of the containers of the cargo using \textit{Qbsolv} with $n=35$, $N=20$, with parameters of Tables \ref{Table:ContainerData} and \ref{Table:OtherData} and constraints \textit{PL} and \textit{CL}. Total resulting mass of $28075$ kg and a center or gravity equal to $-0.83$. The blue vertical lines represent the bounds (respectively $-4$ and $8$) while the green line is the target center of gravity ($4$).}
\includegraphics[width=0.7\textwidth]{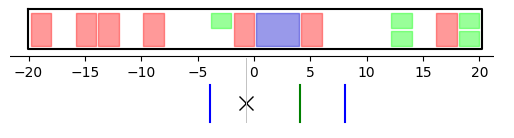}	
\label{Fig:QbsolvPLCLDisp}

	\label{Fig:QbsolvPLCL}
\end{figure*}

\begin{figure*}[!htb]
	\centering
	\caption{This figure shows the profile of the shear curve obtained using \textit{QBsolv} with $n=35$, $N=20$, with parameters of Tables \ref{Table:ContainerData} and \ref{Table:OtherData} and constraints \textit{PL}, \textit{CL} and \textit{SL}. The orange line is the upper shear limit, the blue profile is the shear curve of one solution. }
\includegraphics[width=0.55\textwidth]{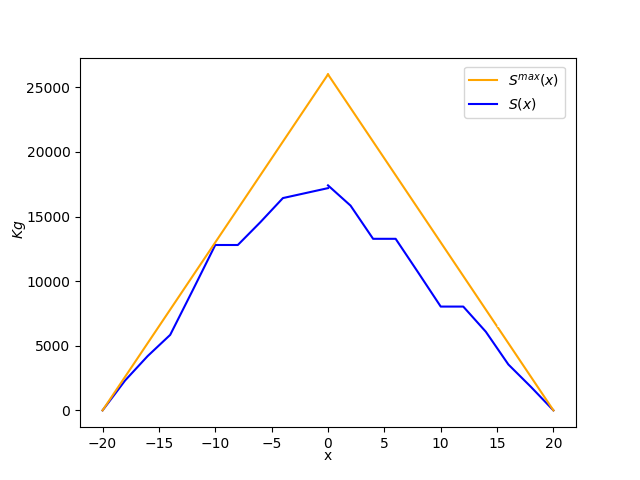}
\label{Fig:QbsolvPLCLSLShear} 
	
	\label{Fig:QbsolvPLCLSL}
\end{figure*}

		\begin{itemize}
			\item \textit{Case PL}. To benchmark the model we first tested the problem considering only the position assignment constraint and the capacity limit imposed to the cargo. The result of one the simulations is shown in Figure \ref{Fig:QbsolvPL}.
			\item \textit{Case PL+CL}. In this benchmark the solution have to be such that all the containers are set in valid positions without exceeding the capacity limit (constraint \textit{PL}) while keeping the center of gravity as close to the target as possible (constraint \textit{CL}). Figure \ref{Fig:QbsolvPLCL} shows one of the obtained solution with the set of constraints \textit{PL} and \textit{CL}.
			\item \textit{Case PL+CL+SL}. The goal of this benchmark was to find a solution in compliance with the constraints \textit{PL}, \textit{CS} and \textit{SL}. The results can be appreciated in Figure \ref{Fig:QbsolvPLCLSL}.
		\end{itemize}
		Tables \ref{Table:NumberOfVariables} shows the number of position assignment variables and slack variables used by each example.

		\subsection{Classical solver - Benchmarking}
		\label{Sub:500}
\begin{figure*}[!htb]
\centering
		\caption{Comparison of the distribution of center of gravity of solution between the cases PL and PL+CL.} 
\label{Fig:CoGDistribution}
	\begin{subfigure}[b]{0.48\textwidth}\centering
		\includegraphics[width=\textwidth]{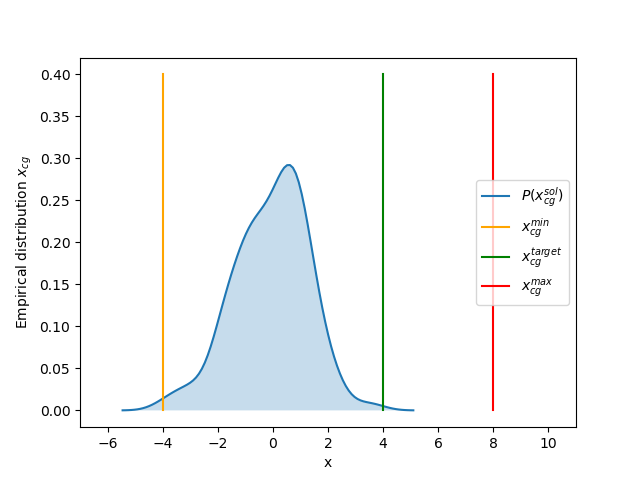}
		\caption{Distribution of the center of gravity (over multiple solutions) using the constraint PL.}
	\end{subfigure}	
	\begin{subfigure}[b]{0.48\textwidth}\centering
		\includegraphics[width=\textwidth]{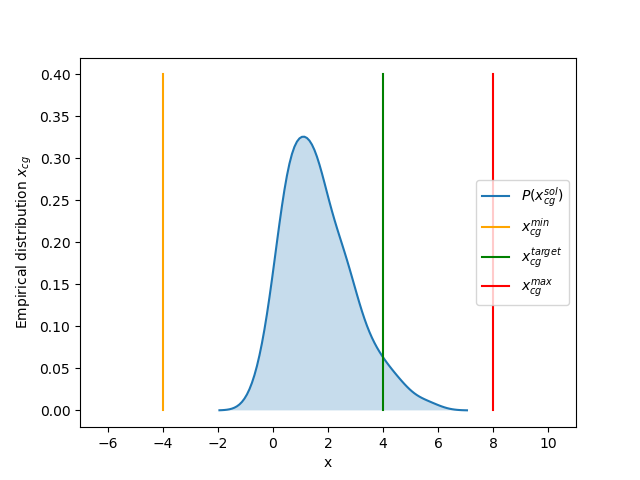}
		\caption{Distribution of the center of gravity (over multiple solutions) using the constraints PL and CL.}
	\end{subfigure}

\end{figure*}

			In order to better analyze the performance of the model introduced in Section \ref{Sec:DetailedExplanation}, we made $500$ runs for each of the set of constraints (PL, PL+CL, PL+CL+SL), keeping the parameters and input values constant through the runs. 
			\begin{itemize}
				\item{\textbf{Average time to solution}}. Average times to solutions are presented in table \ref{Table:Times}. When constraints are added the time to solution is higher due to the additional slack variables.
				\item{\textbf{Success rate comparison}}. Table \ref{Table:SuccessRate} shows the percentage of success rate for the different condition between the 3 main cases (PL, PL+CL, PL+CL+SL). For the Payload condition (PL) we consider it successful if the no-overlap, no-duplication and maximum weight conditions are all satisfied. For the Center of gravity condition (CL) we consider it satisfied if the center of gravity of the solution is between $x_{cg}^{min}$ and $x_{cg}^{max}$. Finally we consider the shear limit (SL) satisfied if the shear limit upper bound is respected in every position of the aircraft.				
				\item{\textbf{Center of gravity}}. in table \ref{Table:SuccessRate} we see that the center of gravity condition is almost always respected even if the constraint is not active. However, a detailed analysis of the results showed that activating the CL constraints did improve the center of gravity position. Figure \ref{Fig:CoGDistribution} shows the comparison of the distribution of center of gravity of the solution between the cases PL and PL+CL. As we can see, the second case provides a better distribution around the target, where both the left and right bounds are respected.
				\item{\textbf{Shear limit}}. When activating the SL constraint the improvement seems less detectable. Table \ref{Table:NumberOfInvalidShear} shows the comparison between the 3 main cases (PL, PL+CL, PL+CL+SL) with respect to the number of positions which led to incorrect values for the shear force. The amount of invalid positions decreases only slightly (this could be due to an imprecise calibration of the SL constraint or to conflicts between the center of gravity condition and the shear limit one, since both act on the disposition of the containers. Further analysis is needed for this case).
			\end{itemize}

	\subsection{D-Wave 2000Q}\label{Sub:DWave}
	When running experiments on the D-Wave 2000Q we used only the payload constraints (PL) and a problem with $n=6$, $N=4$. The amount of slack variables $v_k$ is automatically determined, as explained in section \ref{Sec:DetailedExplanation}. The total number of variables is $48$ (of which $24$ are slack variables).
	In our tests this was the empirical limit for the mapping to the \textit{ChimeraGraph}. We were not able to add the other constraints due to the additional slack variables that would be required.
	
	The problem defined by Table \ref{Table:ContainerDataDWave1} with maximum capacity equal to $8000$ kg has been submitted $100$ times to the D-Wave 2000Q. We compared the results with $100$ runs of the \textit{QBsolv} in terms of computational time and mean/max loaded weight. The results are presented in Table \ref{Table:Dwave500} and \ref{Table:Dwave500W}. \textit{QBsolv} was always able to reach feasible solution, and reached the optimal solution $33\%$ of the times. The D-Wave 2000Q was more noisy, hitting feasible solutions only  $80\%$ of the times and loading on average less weight on the aircraft.

	\section{Conclusion and further developments}
	
	In conclusion \textit{QBsolv} proved to be an efficient tool to validate the model, matching efficiently the exact solver for problem of small dimension. However, for problem of medium/big dimension it is difficult to estimate how far the solutions are from the optimal one. Since the modelling was completed successfully we believe that the only main points where the model could be further improved is a fine tuning of the penalties scaling and of the amount of slack variables used. The optimal solution would be an adaptive and automated selection of penalties and slack variables based on the input data that is submitted to the solver. For the solution through a Quantum Annealer a further development is to test the model on the next generation of D-Wave annealers (based on the Pegasus architecture) that will be released next year (2020).

	\onecolumngrid
\newpage

\section{APPENDIX: TABLES}

\begin{table*}[!htb]
	\captionof{table}{Parameter values used for testing the local solver.}
	\label{Table:OtherData}		
	\begin{tabular}{|c|c|l|}
		\hline
		Symbol & Value & Description\\
		& [Unit of measure]  & \\
		\hline
		N & 20 [1] & Number of available positions\\
		n & 35 [1] & Number of available containers\\
		L & 40 [m] & Length of the payload area on the aircraft\\
		$W_{max}$ & 40000 [Kg] & Maximum payload capacity of the aircraft in kg\\
		$W_{e}$ & 120000 [Kg] & Mass of the aircraft without payload in kg (ready for flight)\\
		$S_{max}$ & 26000 [N] & Max shear force\\
		$x_{cg}^{min}$ & -0,1$\cdot$ L & Minimum allowed centre of gravity position of the aircraft (forward limit)\\						
		$x_{cg}^{max}$ & 0,2$\cdot$ L & Maximum allowed centre of gravity position of the aircraft (aft limit)\\												
		$x_{cg}^{t}$ & 0,1$\cdot$ L & Target centre of gravity position of the aircraft\\								
		\hline
	\end{tabular}
\end{table*}

\begin{table*}[!htb]
	\captionof{table}{Container data used for testing the local solver.}
	\label{Table:ContainerData}		
	\begin{tabular}{|c|c|c|}
		\hline
		Container & Container & Container\\
		ID & type & mass (kg)\\
		\hline
			1 & 1 & 2134\\
			2 & 1 & 3455\\
			3 & 1 & 1866\\
			4 & 1 & 1699\\
			5 & 1 & 3500\\
			6 & 1 & 3332\\
			7 & 1 & 2578\\
			8 & 1 & 2315\\
			9 & 1 & 1888\\
			10 & 1 & 1786\\
			11 & 1 & 3277\\
			12 & 1 & 2987\\
			13 & 1 & 2534\\
			14 & 1 & 2111\\
			15 & 1 & 2607\\
			16 & 1 & 1566\\
			17 & 1 & 1765\\
			18 & 1 & 1946\\
			19 & 1 & 1732\\
			20 & 1 & 1641\\
			21 & 2 & 1800\\
			22 & 2 & 986\\
			23 & 2 & 873\\
			24 & 2 & 1764\\
			25 & 2 & 1239\\
			26 & 2 & 1487\\
			27 & 2 & 769\\
			28 & 2 & 836\\
			29 & 2 & 659\\
			30 & 2 & 765\\
			31 & 3 & 3132\\
			32 & 3 & 3530\\
			33 & 3 & 3892\\
			34 & 3 & 3464\\
			35 & 3 & 3282\\						
		\hline
	\end{tabular}
\end{table*}

\begin{table*}[!htb]
	\captionof{table}{Number of slack variables used in the examples.}
	\label{Table:NumberOfVariables}		
	\begin{tabular}{|c|c|c|}
		\hline
		Case & Pos. assignment & Slack\\
		& variables       & variables \\
		\hline
		PL       & 700 & 71\\
		PL+CL    & 700 & 118\\
		PL+CL+SL & 700 & 386\\			
		\hline
	\end{tabular}
\end{table*}

\begin{table*}[!htb]
	\captionof{table}{Success rate for the different aircraft loading conditions for the different set of constraints}
	\label{Table:SuccessRate}		
	\begin{tabular}{|c|c|c|c|}
		\hline
		Case & \% valid PL & \% valid CL & \% valid SL\\
		\hline
		PL       & 94.4\% & 99.6\% & 58.4\%  \\
		PL+CL    & 98.0\% & 100\% & 63.0\%  \\
		PL+CL+SL & 97.2\% & 100\% & 65.6\%  \\			
		\hline
	\end{tabular}
\end{table*}

\begin{table*}[!htb]
	\captionof{table}{Mean time to solution over 500 tests }
	\label{Table:Times}		
	\begin{tabular}{|c|c|}
		\hline
		Case & Time (s)\\
		\hline
		PL       & 32.9 s  \\
		PL+CL    & 35.5 s  \\
		PL+CL+SL & 38.2 s  \\			
		\hline
	\end{tabular}
\end{table*}

\begin{table*}[!htb]
	\captionof{table}{Shear force errors percentages (0 errors means that all the positions respect the shear limits, 1 error means that one position broke the shear limit and so on).}
	\label{Table:NumberOfInvalidShear}		
	\begin{tabular}{|c|c|c|c|c|}
		\hline
		Case & \% (errors=0) & \% (errors=1) & \% (errors=2) & \% (errors$\geq$3 )\\
		\hline
		PL       & 58.4\% & 23.6\% & 8.6\% & 9.4\% \\
		PL+CL    & 63.0\% & 21.6\% & 10.4\% & 5.0\% \\
		PL+CL+SL & 65.6\% & 24.2\% & 4.8\% & 5.4\% \\			
		\hline
	\end{tabular}
\end{table*}

\begin{table*}[!htb]	
	\captionof{table}{Container data for the experiment on the D-Wave 2000Q.}
	\label{Table:ContainerDataDWave1}		
	\begin{tabular}{|c|c|c|}
		\hline
		Container & Container & Container\\
		ID & type & mass (kg)\\
		\hline
		1 & 1 & 2134\\
		2 & 1 & 3455\\
		3 & 1 & 1866\\
		4 & 1 & 1699\\
		5 & 1 & 3500\\
		6 & 1 & 3332\\					
		\hline
	\end{tabular}
\end{table*}

\begin{table*}[!htb]
	\captionof{table}{Average time to solution, feasibility and optimality for \textit{QBsolv} and D-Wave 2000Q. The connection delay is not included (only computation time).}
	\label{Table:Dwave500}		
	\begin{tabular}{|c|c|c|c|}
		\hline
		Solver & Time (s) & \% Feasible sol. & \% Optimal sol. \\
		\hline
		exact     & 104.3 & - & - \\
		qbsolv    & 0.161 & 100\% & 33\% \\
		dwave     & 0.072 & 80\% & 4\%
	\\			
		\hline
	\end{tabular}
\end{table*}

\begin{table*}[!htb]
	\captionof{table}{Max and Mean solution weight for \textit{QBsolv} and D-Wave 2000Q}
	\label{Table:Dwave500W}		
	\begin{tabular}{|c|c|c|}
		\hline
		Solver & Max Sol. weight &  Mean Sol. weight\\
		\hline
		exact   &  7500 & - \\
		qbsolv  &  7500 & 7394.8\\
		dwave   &  7500 & 4304.0
		\\			
		\hline
	\end{tabular}
\end{table*}
	\onecolumngrid
\newpage
\ \\ 
\section{APPENDIX: IMAGES}

\end{document}